\documentclass[a4paper,11pt]{article}
\usepackage{pos}

\title{Searching for Yang-Lee zeros in O($N$) models}

\author*{Felipe Attanasio}
\author{Marc Bauer}
\author{Lukas Kades}
\author{Jan M. Pawlowski}

\affiliation{Institute for Theoretical Physics, Universit\"{a}t Heidelberg,\\
	Philosophenweg 16, D-69120, Germany}
	
\emailAdd{pyfelipe@thphys.uni-heidelberg.de}
\emailAdd{marc.bauer@thphys.uni-heidelberg.de}
\emailAdd{kades@thphys.uni-heidelberg.de}
\emailAdd{j.pawlowski@thphys.uni-heidelberg.de}

\abstract{
Near the second order phase transition point, QCD with two flavours of massless quarks can be approximated by an O($4$) model, where a symmetry breaking external field $H$ can be added to play the role of quark mass.
The Lee-Yang theorem states that the equation of state in this model has a branch cut along the imaginary $H$ axis for $|Im[H]|>H_c$, where $H_c$ indicates a second order critical point.
This point, known as Lee-Yang edge singularity, is of importance to the thermodynamics of the system.
We report here on ongoing work to determine the location of $H_c$ via complex Langevin simulations.
}

\FullConference{%
 The 38th International Symposium on Lattice Field Theory, LATTICE2021
  26th-30th July, 2021
  Zoom/Gather@Massachusetts Institute of Technology
}

\newcommand{\OO}{\mathcal{O}}

\begin{document}
\maketitle

\section{Introduction}
Around the chiral phase transition the matter dynamics of QCD can be well-approximated by a theory of massless pions and scalar quark condensate.
In this low energy regime, nucleons and heavier hadrons play no role, and the model is described by an $O(4)$ scalar field with quartic self-interactions.
In this model, an external source $h_0$ coupled to the quark condensate plays the role of chemical potential.
The breaking of chiral symmetry, signalled by a vacuum expectation value of the quark condensate, is controlled by the sign and strength of the mass parameter.

The Yang-Lee theorem~\cite{PhysRev.87.404,PhysRev.87.410} states that O($N$) models have branch cuts for purely imaginary values of $h_0$, and the cuts end at second order critical points, known as Lee-Yang edge singularities.
These singularities provide an upper bound on the radius of convergence of a Taylor expansion in $h_0$ around the origin.
Analogously, in QCD this would correspond to an expansion in the chemical potential $\mu$ around the chiral phase transition.
Thus, knowing the location of the edge singularity in the O($N$) case provides information on how far in the chemical potential a Taylor expansion in QCD can be trusted.
An earlier investigation on this topic, using random matrix models, can be found in ref.~\cite{Stephanov:2006dn}.
Recent works on this topic have been carried out using Taylor expansions~\cite{Giordano:2019slo}, lattice QCD results for non-universal parameters~\cite{Mukherjee:2019eou}, reweighting~\cite{Giordano:2019gev}, Functional Renormalisation Group methods~\cite{connelly_universal_2020}, Pad\'e resummations~\cite{Basar:2021hdf}, as well as studies in QCD via imaginary chemical potential~\cite{Nicotra:2021ijp,Singh:2021pog}.

The complex external magnetic field, used to probe the analytic structure of the O($N$) model, makes the action complex, leading to a sign problem (or complex phase problem).
In order to circumvent it, we make use of the complex Langevin (CL) method~\cite{Klauder:1983nn,Klauder:1983zm,Klauder:1983sp,Parisi:1984cs}, an extension of stochastic quantisation~\cite{Parisi:1980ys}.
This method has been used over the past decades with success in different contexts, such as ultracold atoms~\cite{Hayata:2014kra,Rammelmuller:2018hnk}, QCD~\cite{Sexty:2013ica,Aarts:2016qrv,Attanasio:2018rtq}, and superstring inspired models~\cite{Nishimura:2019qal}.
For a recent reviews of applications of complex Langevin, see~\cite{Berger:2019odf,Attanasio:2020spv} and references therein.

\section{Computational method}
The action describing the O($N$) field in $3$ spatial dimensions is given by
\begin{equation}
	S = \int d^3 x \left[ \frac{1}{2} \partial_\mu \varphi_i \partial_\mu \varphi_i + \frac{m_0^2}{2} \varphi_i \varphi_i + \frac{\lambda_0}{4!} (\varphi_i \varphi_i)^2 + h_0 \varphi_1 \right]\,,
\end{equation}
where $m_0 \in \mathbb{R}$ is the mass, $\lambda_0 \in \mathbb{R}$ is quartic coupling, $h_0 = h_R + i h_I \in \mathbb{C}$ represents the (complex) external magnetic field, and $1 \leq i \leq N$.
Due to the presence of $h_0$ the action is complex, and thus cannot be simulated using standard Monte Carlo methods that rely on importance sampling.

We employ the complex Langevin method to generate field configurations distributed according to the weight $e^{-S}$.
In this method the fields are augmented by a fictitious time dimension $\tau$, known as Langevin time, over which they evolve according to the Langevin equation
\begin{equation}
	\frac{d \varphi_i(x, \tau)}{d \tau} = -\frac{\delta S[\varphi_i(x,\tau)]}{\delta \varphi_i} + \eta_i(x,\tau)\,,
\end{equation}
with $\eta_i(x, \tau)$ being white noise fields
\begin{equation}
			\langle \eta \rangle = 0 \,, \quad \langle \eta_i(x, \tau) \eta_j(y, \tau') \rangle = 2 \delta_{ij} \, \delta(x-y) \delta(\tau-\tau')\,,
\end{equation}
and the $\langle \cdots \rangle$ represent ensemble averages.
Quantum expectation values are computed via simple averages over the Langevin time, after the system reaches its equilibrium state.

Due to the complexity of the action, all real fields are also given an imaginary part.
In this enlarged manifold where the fields are defined complex Langevin is not guaranteed to converge to the correct result.
Extensive discussions on criteria for correct convergence can be found in refs.~\cite{aarts_complex_2011, nagata_argument_2016, Scherzer:2018hid, scherzer_controlling_2020,seiler_complex_2020}.

Our numerical simulations use the Euler-Maruyama discretisation for the Langevin equation
\begin{equation}
	\varphi_i(x, \tau + \epsilon) = \varphi_t(x, \tau) - \epsilon \frac{\delta S[\varphi_i(x,\tau)]}{\delta \varphi_i} + \sqrt{\epsilon} \, \eta_i(x,\tau)
\end{equation}
with the Langevin step size $\epsilon$ being changed adaptively~\cite{aarts_adaptive_2010} to avoid runaway trajectories.
A study of implicit schemes to numerical evolve the complex Langevin equation is found in ref.~\cite{Alvestad:2021hsi}.

The Lee-Yang theorem states that the magnetic equation of state in the symmetric phase exhibits branch cuts that end at the edge singularities.
Therefore, we are interested in the average magnetisation per site
\begin{equation}
	M = \frac{1}{V}\frac{\partial}{\partial h_0} \ln Z = \frac{1}{V} \left\langle \Phi_1 \right\rangle\,, \quad \Phi_1 = \int d^3x \, \varphi_1(x)  
\end{equation}
and its susceptibility
\begin{equation}
	\chi = \frac{1}{V} \frac{\partial^2}{\partial h_0^2} \ln Z = \frac{1}{V} \left[ \left\langle \Phi_1^2 \right\rangle - \langle \Phi_1 \rangle^2 \right]\,.
\end{equation}
The magnetisation should exhibit a discontinuity when $h_R$ goes across the branch cut at $h_R=0$ and $|h_I| > h_c$, whereas the susceptibility should peak (due to finite lattice volume) for $|h_I| \approx h_c$.

\section{Results}
Our studies started with a toy model: an one site model with O($2$) symmetry.
A very similar case, with a single field instead of two, has been considered in ref.~\cite{Salcedo:2018uop}.
There, a failure of CL has been observed.

Afterwards, we have moved to a three dimensional theory.
Since in the thermodynamic limit the Lee-Yang theorem states that branch cuts exist, the analytic structure of the partition function in this model is qualitatively different from the $0$-dimensional case.
It is then interesting to see how complex Langevin fares in this scenario.

\subsection{Single site O(\texorpdfstring{$2$}~) model}
We begin with an analytically solvable one site model at $N=2$, given by the action
\begin{equation}
		S = \frac{m_0^2}{2}\left(x^2 + y^2\right) + \frac{\lambda_0}{4}\left(x^2 + y^2\right)^2 - h_0 x\,.
\end{equation}
Its partition function and magnetisation are, respectively, given by
\begin{equation}
		Z = 2 \pi \int_0^\infty r dr \, I_0(h_0 r) \exp\left[ -\frac{m_0^2 r^2}{2} - \frac{\lambda_0 r^4}{4} \right]\,,
\end{equation}
and
\begin{equation}
		M = \langle x \rangle
		= \frac{\partial \ln Z}{\partial h_0}
		= \frac{2 \pi}{Z} \int_0^\infty r^2 dr \, I_1(h_0 r) \exp\left[ -\frac{m_0^2 r^2}{2} - \frac{\lambda_0 r^4}{4} \right]\,,
\end{equation}
where $I_n(x)$ are the modified Bessel functions of the first kind.

\begin{figure}
	\begin{tabular}{cc}
	\includegraphics[width=0.45\textwidth]{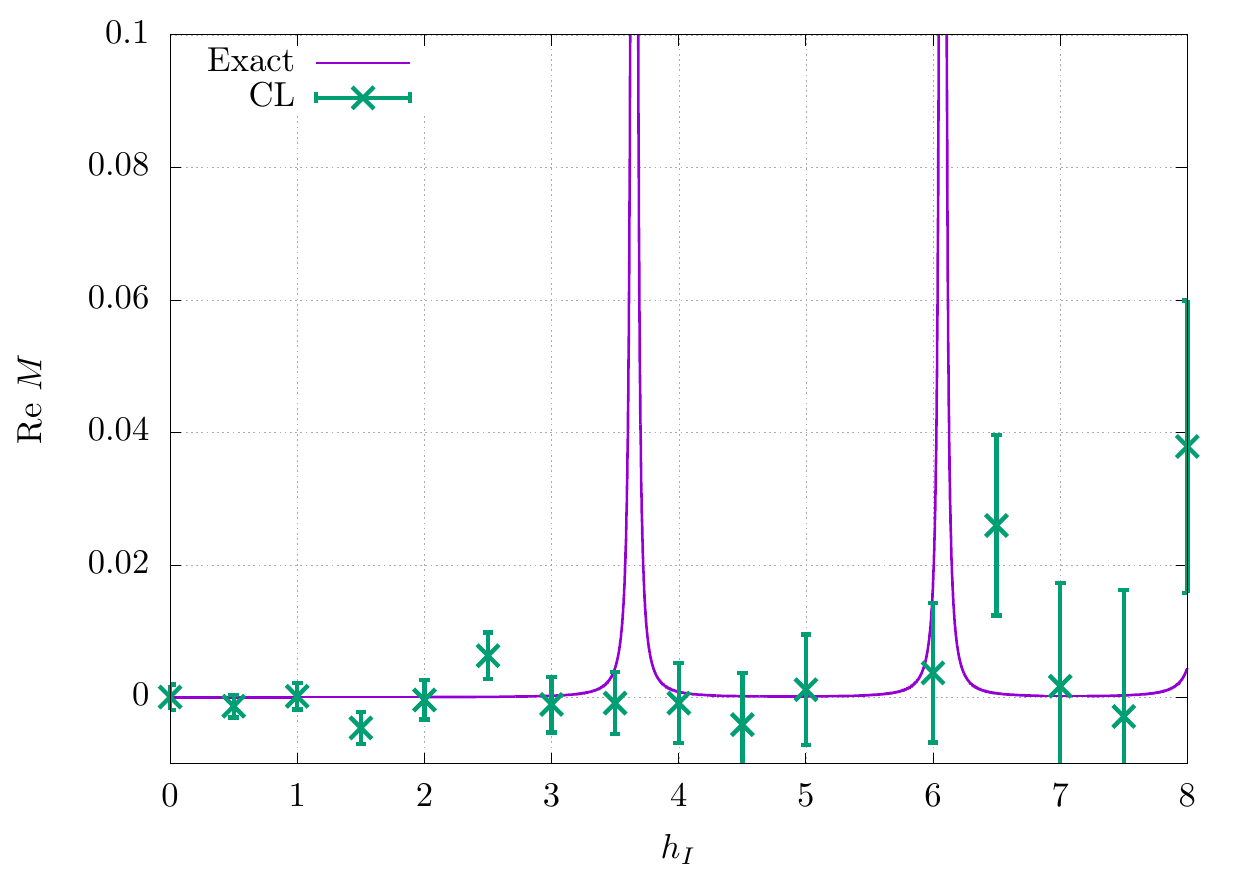} &
	\includegraphics[width=0.45\textwidth]{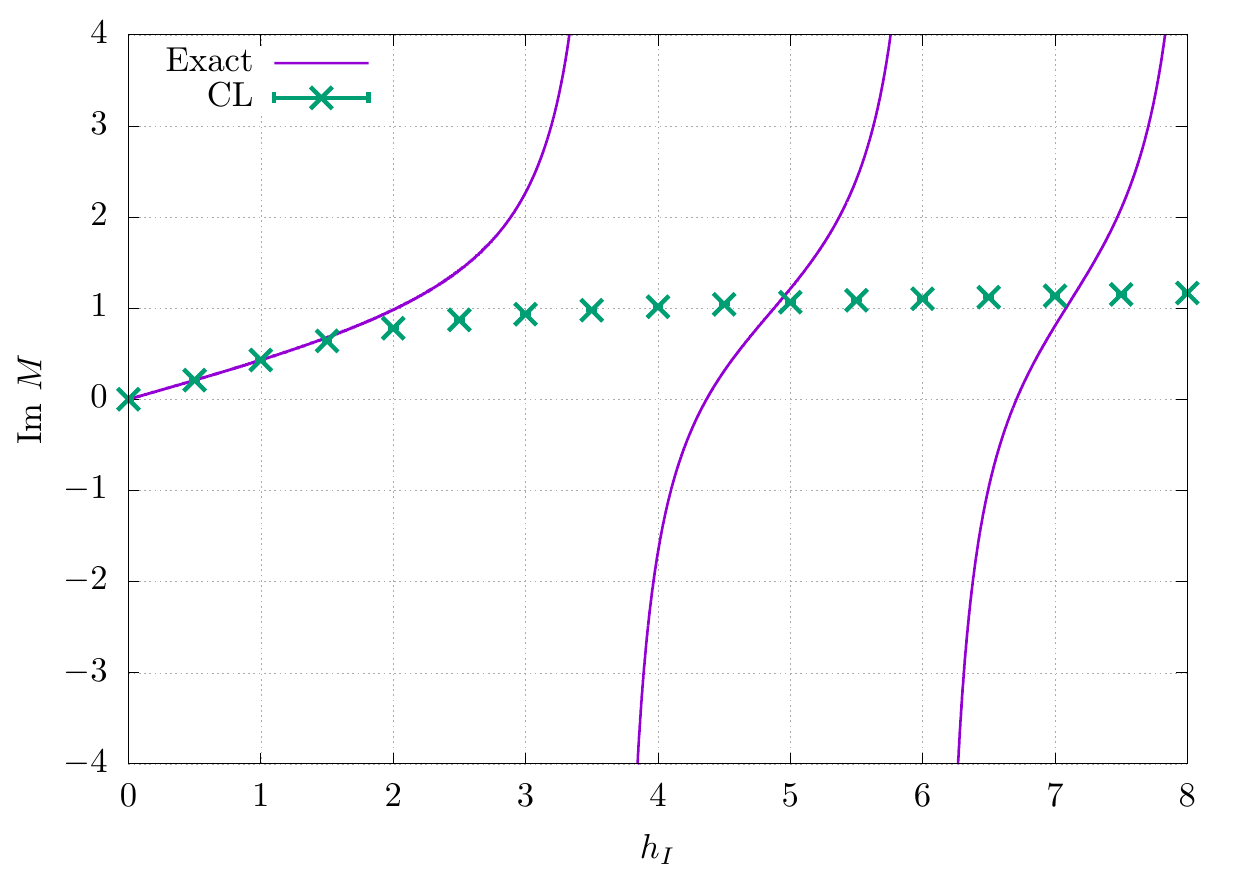}
	\end{tabular}
	\caption{\label{fig.sitemodel}Real and imaginary parts of the magnetisation of the single site O($2$) model as function of $h_i$.
		The continuous lines represent the exact solutions, while the points resulted from complex Langevin simulations.
		The divergences correpond to zeros of the partition function, which the complex Langevin fails to capture.
		}
\end{figure}

As it is shown in fig.~\ref{fig.sitemodel}, the complex Langevin simulations fail to capture the position of the edge, and all subsequent, singularities.
This is similar to what has been observed in studies with Random Matrix theory~\cite{Bloch:2017sex} and models with singular Langevin drifts~\cite{nishimura_new_2015}.
It is worth noting that away from the thermodynamic limit the branch cuts predicted by the Yang-Lee theorem become a series of poles that coalesce as $V\to\infty$.

\subsection{Three dimensional field theory}
\begin{figure}
	\begin{tabular}{cc}
	\includegraphics[width=0.45\textwidth]{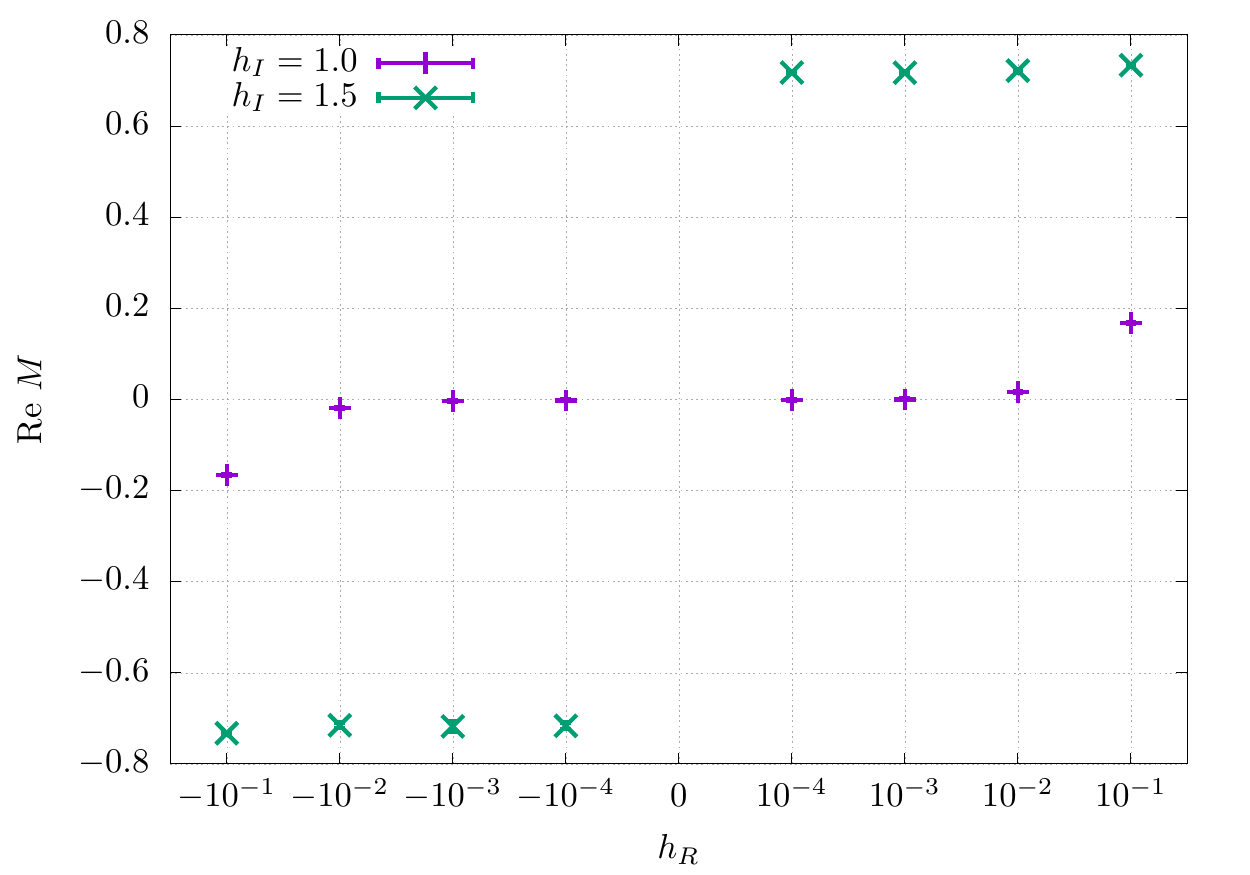} &
	\includegraphics[width=0.45\textwidth]{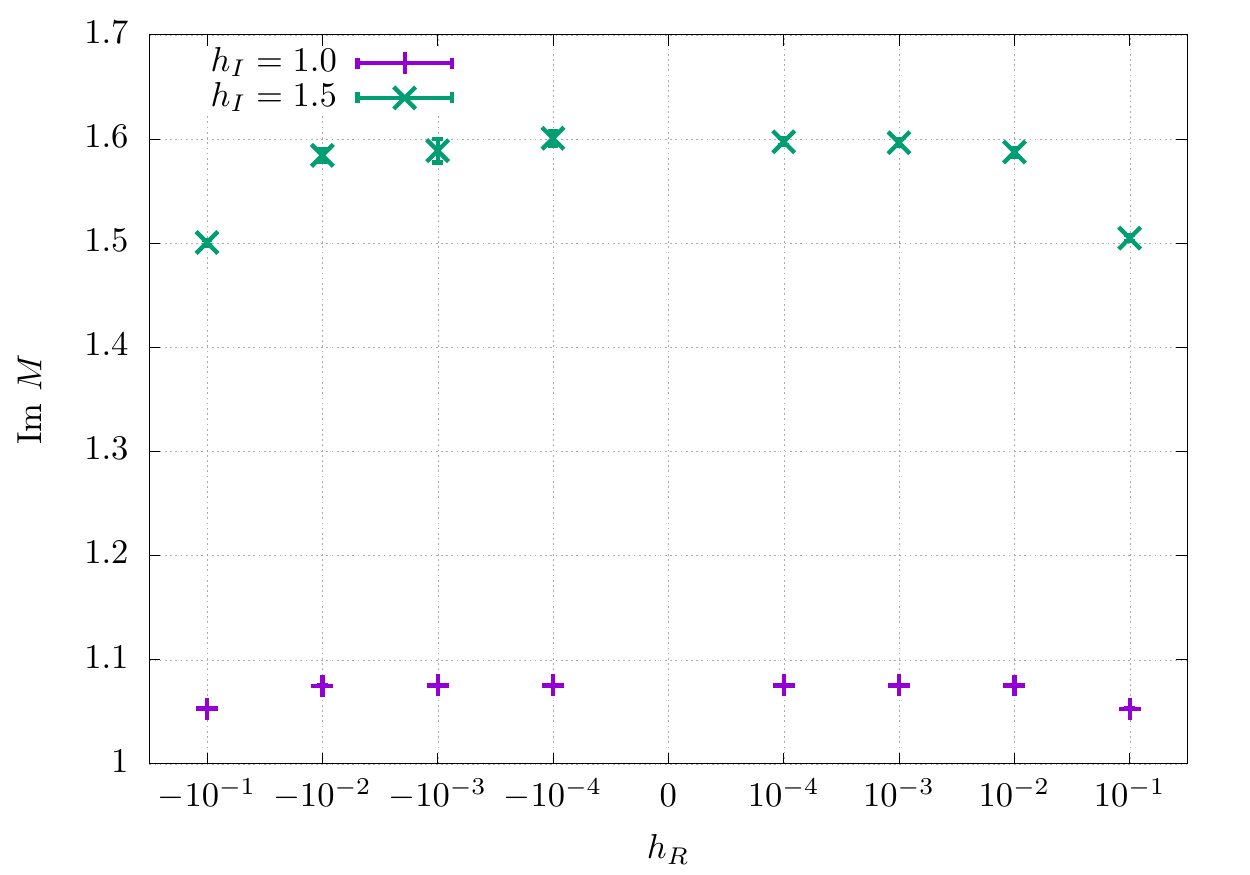}
	\end{tabular}
	\caption{\label{fig.hre.scan}
		Real (left) and imaginary (right) parts of the average magnetisation as functions of the real part of the external magnetic field.
		Note that the horizontal axis is in units of powers of $10$, except at $0$.
		The graph on the left shows the two expected behaviours for the magnetisation: for $h_I < h_c$ it is continuous across $h_R=0$, and it displays a discontinuity for $h_I > h_c$.
}
\end{figure}

We have performed our simulations with parameters $m_0^2 = \lambda_0 = 1$ and $N=2$ for various values of $h_0$.
Preliminary, small volume, simulations have been carried out in $V=8^3$ in order to verify the (dis)continuity of the magnetisation for $h_I$ smaller (larger) than $h_c$ as function of $h_R$, by choosing two representative values of $h_I$.
The results are shown in figure~\ref{fig.hre.scan}, and demonstrate the expected discontinuity for $h_I > h_c$.

\begin{figure}
	\begin{tabular}{cc}
	\includegraphics[width=0.48\textwidth]{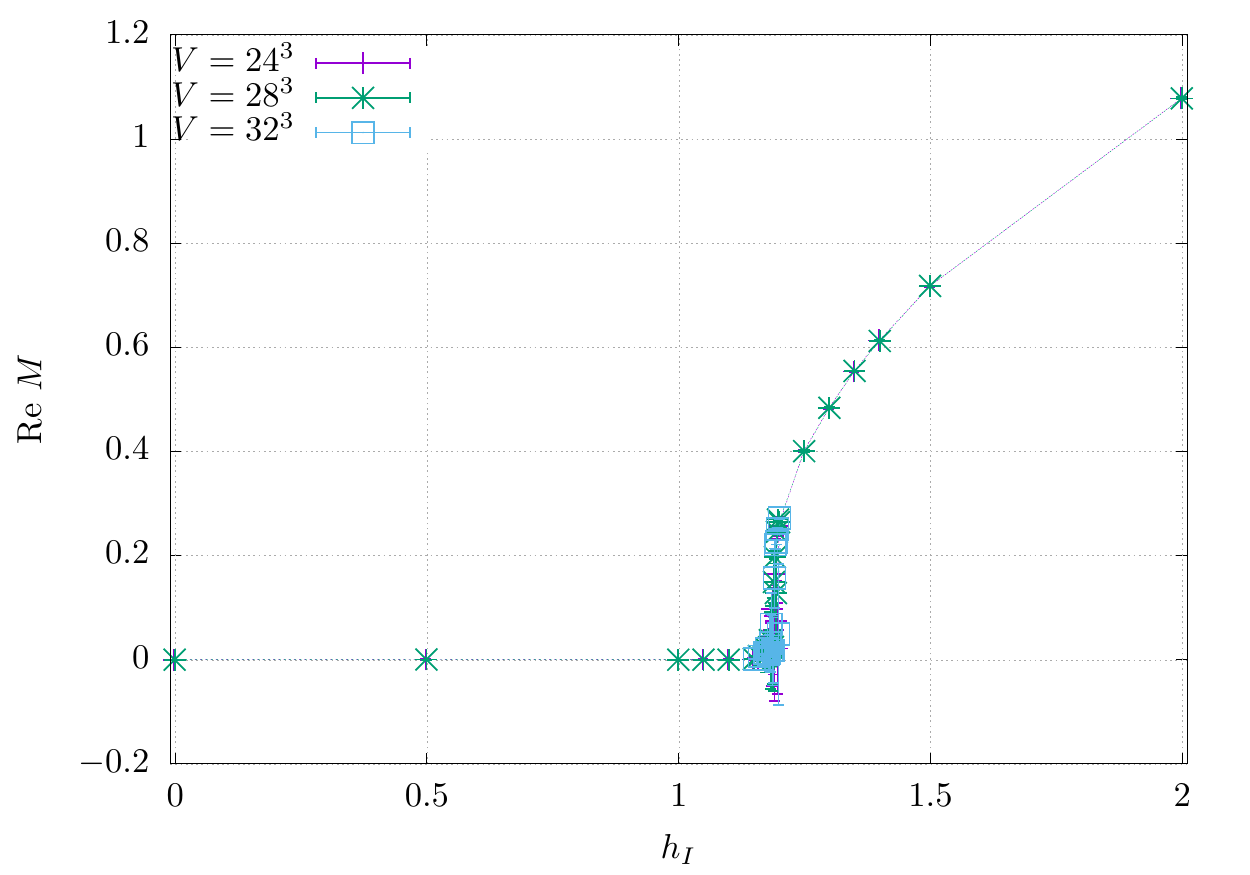} &
	\includegraphics[width=0.48\textwidth]{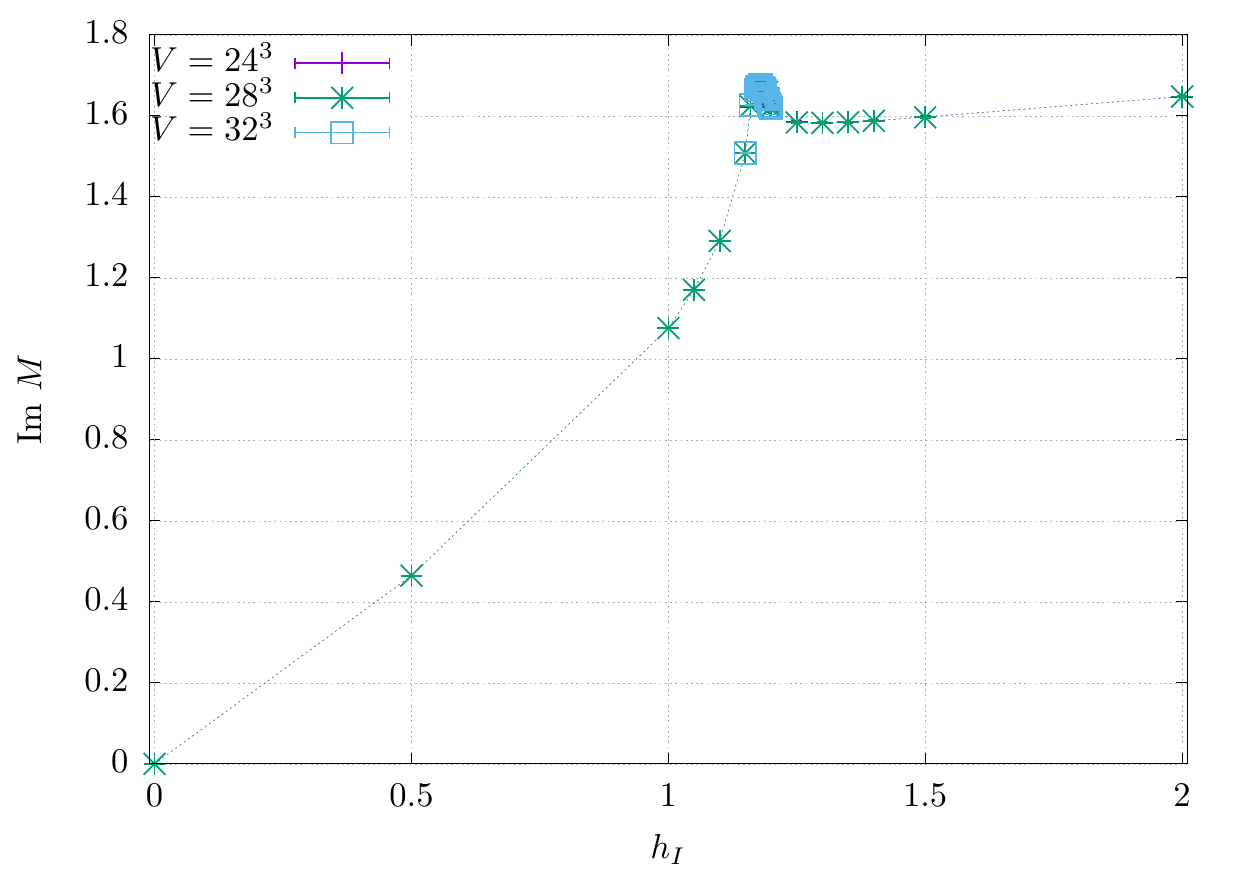}
	\end{tabular}
	\caption{
		\label{fig.magn}
		Real (left) and imaginary (right) parts of the average magnetisation as functions of the imaginary part of the external magnetic field.
		The results for the three different volumes considered here fall on top of each other in the scale used in the plot.
		It is notable that for $h_I \approx 1.185$ there is a kink in the real part of $M$, while around the same value the imaginary part displays a cusp.
		Lines have been added to guide the eye.
		}
\end{figure}

\begin{figure}
	\begin{tabular}{cc}
	\includegraphics[width=0.48\textwidth]{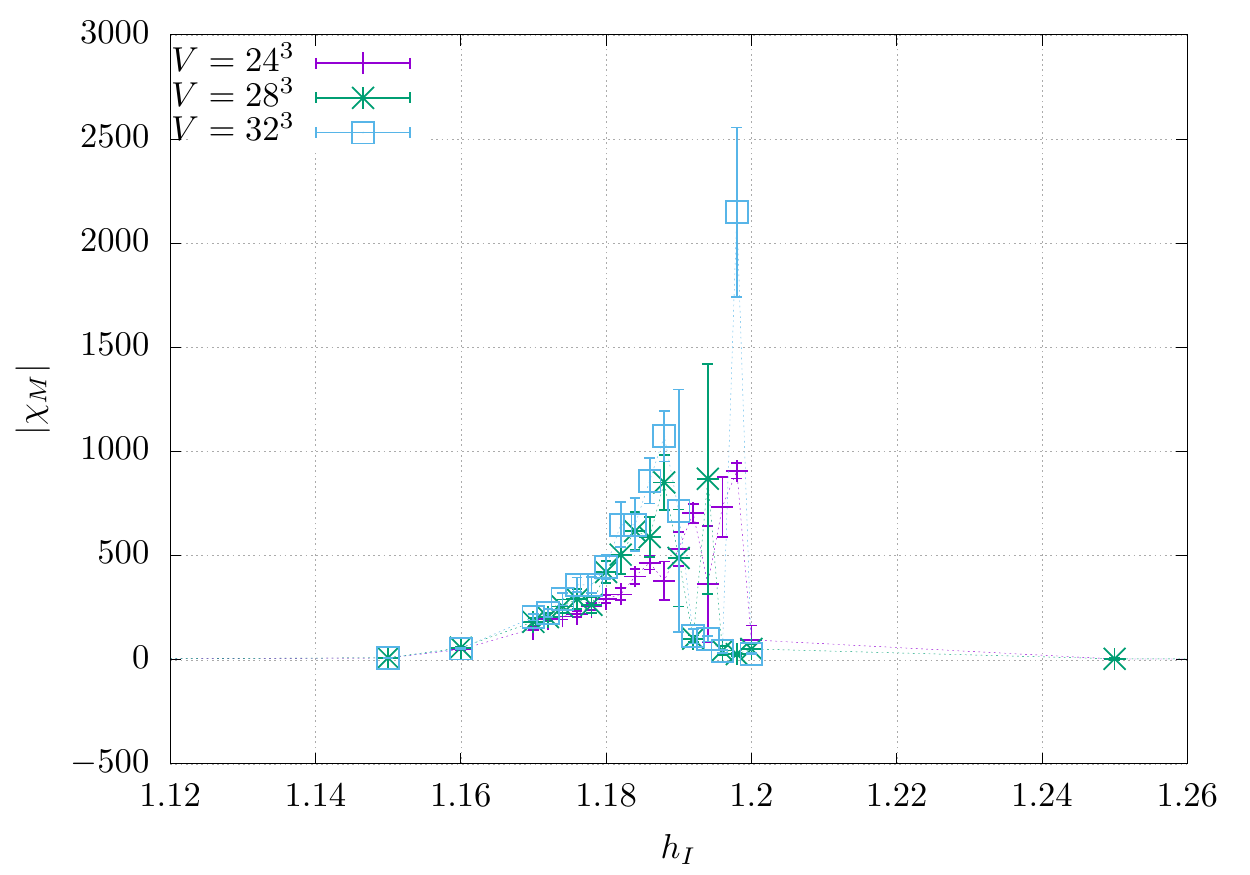} &
	\includegraphics[width=0.48\textwidth]{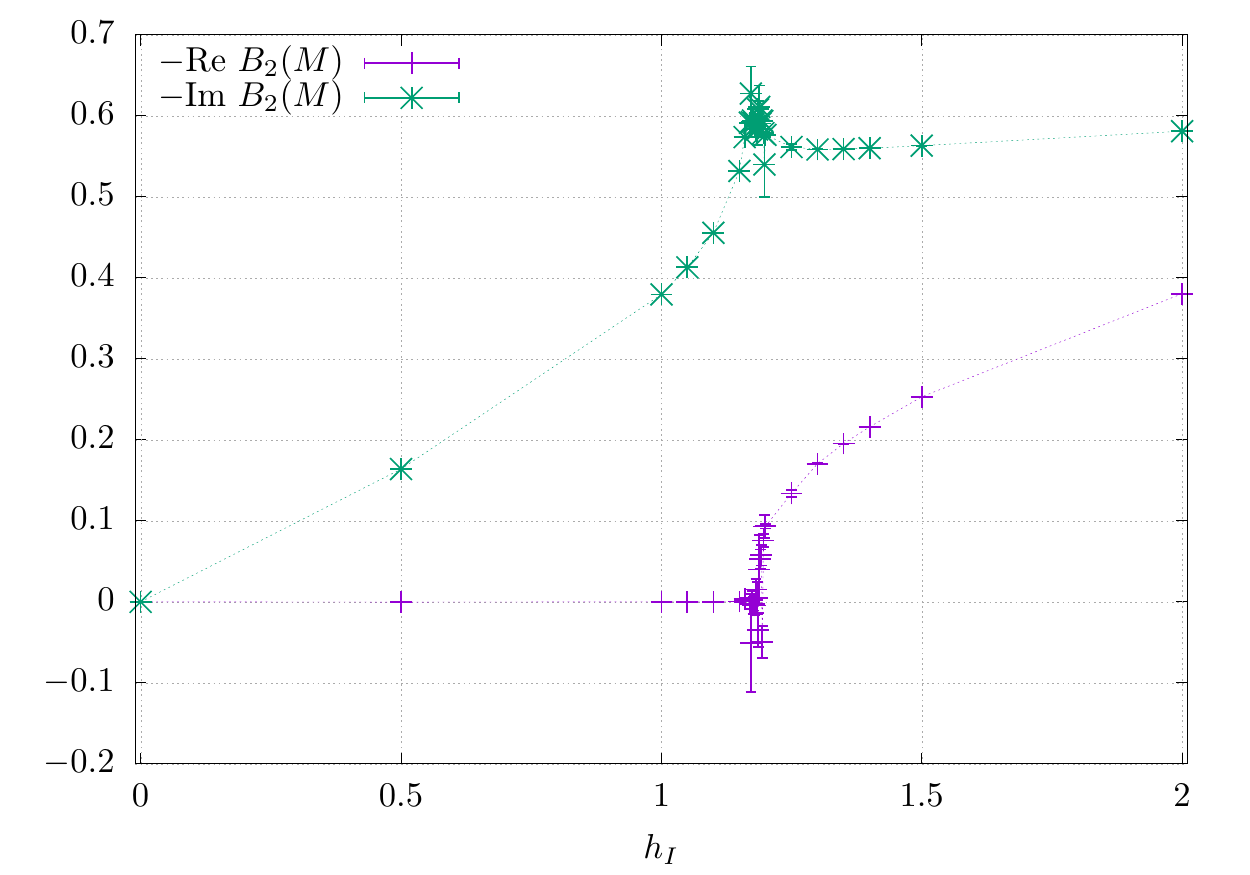}
	\end{tabular}
	\caption{
		\label{fig.susc}
		Left: absolute value of the (complex) magnetic susceptibility as a function of $h_I$ for the three volumes considered.
		The peaks exhibit the typical behaviour of finite volume phase transitions of growing with the volume.
		Right: Negative of the real and imaginary parts of the second order boundary term for the magnetisation, computed for $V=28^3$.
		Similar behaviours have been observed for the other volumes.
		In both plots lines have been added to guide the eye.
		}
\end{figure}

A proper scan for the critical external field along the imaginary axis\footnote{Our simulations have been performed with $h_R = 10^{-4}$, akin to what is usually done in studies of the Ising model.} was then performed at volumes of $24^3$, $28^3$, and $32^3$.
Figure~\ref{fig.magn} presents our results for the magnetisation as function of $h_I$.
The real part resembles a graph of the magnetisation as a function of inverse temperature of the Ising model.
Both plots also exhibit non-differentiable behaviour around a certain value of $h_I$, with the imaginary part having a cusp, rather than a kink.
This is further investigated in figure~\ref{fig.susc} (left) where we present the magnetic susceptibility.
In the usual telltale sign of ``phase transitions'' in finite volumes, the susceptibilities have peaks where the magnetisation exhibited non-differentiable behaviour.

We have also computed the first two complex Langevin boundary terms~\cite{scherzer_controlling_2020} for the magnetisation and present the results for the second order one, $B_2(M)$ in fig.~\ref{fig.susc} (right).
The first order boundary terms vanish within statistical errors.
Since $B_2(M)$ is in general larger than $B_1(M)$ in the cases considered here, the correction formula proposed in ref.~\cite{scherzer_controlling_2020} is not applicable.
Moreover, this observerd hierarchy of boundary terms casts a shadow on the reliability of complex Langevin simulations in this model.
It is also noteworthy that $B_2(M)$ follows the same trend as the magnetisation itself.

\section{Summary and outlook}
We have studied the O($N$) model, subject to a complex external magnetic field, on the lattice via complex Langevin simulations.
Our work has been performed at $N=2$ both a single site model, with analytic solution, and on a three dimensional field theory.

Complex Langevin simulations on the single site model failed to exhibit the correct behaviour, behaving similarly to what has been observed in Random Matrix Theories and models with singular drifts.
In three dimensions, complex Langevin results for the average magnetisation and its susceptibility behave similarly to Ising model simulations.
From analysing the peak of the susceptibility for different volumes it is possible to find the critical value of the external imaginary magnetic field and, therefore, the location of the Lee-Yang edge singularity.
This is very promising.
However, boundary terms are present and thus CL results must be interpreted with care.
This is the subject of ongoing research, both in order to verify the correctness of CL applied to this model and to find the location of the Lee-Yang edge singularity.

\acknowledgments
This work is supported by the Deutsche Forschungsgemeinschaft (DFG, German Research Foundation) under Germany’s Excellence Strategy EXC2181/1 - 390900948 (the Heidelberg STRUCTURES Excellence Cluster) and under the Collaborative Research Centre SFB 1225 (ISOQUANT).

The authors acknowledge support by the state of Baden-W\"urttemberg through bwHPC.

\bibliographystyle{JHEP}
\bibliography{ON}

\end{document}